\documentclass[a4paper,reqno,12pt,draft]{article}
\usepackage{amssymb,euscript,bbold}

\newcommand{\be}{\begin{equation}}
\newcommand{\ee}{\end{equation}}
\newcommand{\ba}{\begin{eqnarray}}
\newcommand{\ea}{\end{eqnarray}}

\newcommand{\ft}{\footnote}

\newcommand{\we}{\wedge}

\begin{document}
\input{epsf}

\begin{flushright}
QMW-PH-00-08
%26th May 1998
\end{flushright}
\begin{flushright}
%{\sf \today}
\end{flushright}
\begin{center}
\Large{\sc Flux, Supersymmetry and $M$ theory on 7-Manifolds.}\\
\bigskip
{\sc B.S. Acharya}\ft{r.acharya@qmw.ac.uk}$^,$
\ft{Work supported by PPARC}\\
and\\
{\sc B. Spence}\ft{b.spence@qmw.ac.uk}\\
\smallskip\large
{\sf Department of Physics,\\
Queen Mary and Westfield College,\\ Mile End Road,\\ London. E1 4NS.}

\end{center}
%\renewcommand{\abstractname}{\sc Abstract}
%\begin{Abstract}
\bigskip
\begin{center}
{\bf {\sc Abstract}}
\end{center}
%\bigskip
%\normalsize
Various aspects of low energy $M$ theory compactified
to four dimensions are considered. If the supersymmetry parameter is parallel
in the unwarped metric, then
supersymmetry requires that the warp factor is trivial, the
background
four-form field strength is zero
and that the internal 7-manifold
has $G_2$ holonomy
(we assume the absence of boundaries and other
impurities). A proposal of Gukov - extended here
to include $M$2-brane domain walls - for the superpotential of the
compactified theory is shown to yield the same result. Finally, we make some
speculative remarks concerning higher derivative corrections and
supersymmetry breaking.

%\end{center}
%\end{abstract}
\newpage

\Large
\noindent
{\bf {\sf 1. Introduction.}}
\normalsize
\bigskip

At low energies, $M$ theory is described by eleven dimensional supergravity
theory. The latter theory was presented in \cite{cjs}.
Over the past few years $M$ theory studies have led to a reconsideration
of results in the supergravity theory.
In the 1980's much attention was
given to Kaluza-Klein compactifications of eleven dimensional supergravity
to four dimensions and we will reconsider some aspects of this work here.
For a review article on Kaluza-Klein supergravity see \cite{KKS}.

The motivation for our work stemmed primarily from certain results concerning
$M$ theory on Calabi-Yau fourfolds \cite{bb,gflux,gvw,greene},
and in particular the
observations due to Becker and Becker that one can include a non-trivial
background expectation value for the four-form field strength in a
supersymmetric fashion. Gukov, Vafa and Witten later proposed a simple
formula for the superpotential of the compactified theory
in such backgrounds,
and showed that the resulting conditions for unbroken supersymmetry are
precisely the conditions obtained by Becker and Becker who analysed
the supersymmetry equations of eleven dimensional supergravity. Gukov
later went on to propose similar formulae for the superpotential of $M$ theory
compactified on various other special holonomy manifolds \cite{guk}.

In this paper we will discuss the relationship between the four-form field
strength and supersymmetry in compactifications of $M$ theory to four
dimensions and the relationship (when it exists) to Gukov's proposed
superpotential. We will begin by analysing supersymmetry conditions in
eleven dimensional supergravity backgrounds with a spacetime metric which
is a warped product of a four-metric and a 7-metric. We will see that
supersymmetry requires the warp factor to be trivial if the supersymmetry
parameter is parallel in the unwarped metric.
This implies that the 7-manifold has
$G_2$-holonomy and that the four-form field vanishes identically\ft{For
trivial warp factor, this is the calculation of Candelas and Raine
\cite{cr}. For completeness,
we extend their calculation to the case of non-zero cosmological constant.
The only solution in this case appears to be the usual supersymmetric
Freund-Rubin ansatz \cite{fr} in which both the four and seven manifolds
admit Killing spinors. ${AdS_4}{\times}{S^7}$ is the standard example}.

We will go on to show how
Gukov's proposed
superpotential for $G_2$-holonomy
compactifications also gives rise to the same
conclusion: that the only non-zero field in a supersymmetric
background is the metric. This is in
contrast to the situation in three dimensions where Becker and Becker
showed that both the warp factor and the four-form field can be non-zero
in a supersymmetric background.

The same analysis can be done for compactifications to three dimensions on
$Spin(7)$ holonomy 8-manifolds and an agreement between Gukov's proposal
for that case and the supergravity analysis is also found. In this case, one
can again turn on a non-trivial warp factor and four-form field strength in a
manner consistent with supersymmetry. The details for $Spin(7)$ holonomy
compactifications will be presented elsewhere \cite{bx}.

In the next section we present the details of the supergravity calculation.
In section three we show that Gukov's proposed superpotential gives the
same result. Finally we end with some discussion concerning higher
derivative $M$ theoretic corrections to our results and speculate on possible
scenarios with low energy supersymmetry breaking. Our conventions are
detailed in an appendix.

There has been a previous attempt in the literature to consider the
relationship between the supergravity equations and Gukov's potential
 \cite{her}. Unfortunately, we do not agree with the results of that paper.

\newpage

\Large
\noindent
{\bf {\sf 2. Calculation.}}
\normalsize
\bigskip

We will consider backgrounds of low energy M theory in which
the spacetime 11-manifold $X$ decomposes into a warped product
of a four-manifold $M^4$ with metric ${g_4}(x)$
and a 7-manifold $M^7$ with metric ${g_7}(y)$. The
simplest case to consider would be when $g_4$ is the Minkowski metric
on ${\bf R^4}$ and $M^7$ is compact in which case we can intepret the
background as a compactification of $M$ theory to four dimensional Minkowski
space. We will take $M^4$ to have Lorentzian signature.
The most general ansatz for the 11-metric is
\be
g_{11}(x,y) =
{{\Delta}}^{-1}(y) (g_{4}(x) + g_{7}(y))
\ee
which describes a warped product metric on ${M^4}{\times}{M^7}$.
We are interested in the constraints
that supersymmetry imposes on such backgrounds, so we require that the
expectation value of the gravitino field vanishes in the background.
With this choice the equations for supersymmetry in the background
are just that
the supersymmetric variation of the gravitino must vanish.
This means  that there exists
a spinor $\eta$ for which
\be
{\nabla}_{M} {\eta} + Z_{M} {\eta} = 0
\ee
where
\be
Z_M = {1 \over 144}({{\Gamma}_M}^{PQRS} - 8{\delta}_M^P {\Gamma}^{QRS})
G_{PQRS}
\ee
The above equations are valid in the extreme low energy limit of $M$ theory
and in principle receive higher derivative corrections. We will
comment on these corrections later. We are also assuming that the
background is free of boundaries, fivebranes
or other `impurities' which can give rise to additional terms
(see, for instance \cite{wit}.)
The above metric
has a group of symmetries
which act on the frame bundle of $X$ and this group
is locally ${SO(3,1)}{\times}{SO(7)}$ where the two factors act obviously on
${M^4}$ and ${M^7}$ respectively.
The most general ansatz for the four-form field strength $G$,
consistent with the spacetime symmetries is:
\ba
G_{\alpha \beta \gamma \delta} = 3m{\epsilon}_{\alpha \beta \gamma \delta} \;\;
,\;\;\;G_{mnpq}  \neq  0  \;\;
\ea
with all other components vanishing. The factor of three is as usual for
convenience and $m$ is a constant.
Our goal is to describe the constraints
on $m$, $G_{mnpq}$ and ${\Delta}(y)$ imposed by supersymmetry.

Consider first the $\mu$ components of $(2)$. After substituting $(1)$ and
$(4)$, writing the connection as the spin connection for
${\Delta}(y)g_{11}(x,y)$  plus terms involving derivatives of the warp factor
(cf. \cite{bb}) we find:
\be
{\nabla}_{\mu} {\eta} + Z_{\mu} {\eta} = 0
\ee
where
\be
Z_{\mu} = {\gamma}_{\mu} {\otimes} \alpha + {\gamma}_5 {\gamma}_{\mu}
{\otimes} ({\beta}_1 + i{\beta}_2 ),
\ee

\be
{\alpha} \equiv {{\Delta}^{3/2} \over 144} G_{mnpq}{\gamma}^{mnpq}
\ee
and
\be
{\beta}_1 \equiv {1 \over 4} {\gamma}^{m}{\partial}_{m}{\log}(\Delta) \;\;
,\;\;\; \beta_2 \equiv -{\Delta}^{3/2} m
\ee
Initially we will assume that
$({M^4},g_{4}(x))$ are such that they admit
parallel (ie covariantly constant) spinors (Minkowski space
being the prime example)
in which case $(2)$ becomes
\be
{Z_{\mu}} \eta = 0
\ee
Since ${\gamma}_{\mu}$ is invertible, we obtain
\be
\alpha \eta = {\gamma}_5 \otimes ({\beta}_1 + i {\beta}_2 ) \eta
\ee

Now consider the $p$-components of $(2)$. Substituting our ansatz we obtain
\be
{\nabla}_p \eta - {1 \over 4} {\gamma}_{p}^{\;n}{\partial}_n \log (\Delta )
\eta +
{\gamma}_{5} \otimes {\gamma}_p \alpha \eta
- {\Delta}^{3 \over 2} ({1 \over 12}{\gamma}_{5} \otimes G_{pqrs}{\gamma}^{qrs} \eta
+ {i \over 2} m
\otimes {\gamma}_m \eta ) = 0
\ee
Contracting this equation with ${\gamma}^p$ and substituting $(10)$ we obtain,
\be
{\gamma}^p {\nabla}_p \eta - {11 \over 4} {\gamma}^p {\nabla}_p \log (\Delta
) \eta + {3i \over 2} m {\Delta}^{3 \over 2} \eta = 0
\ee

Our assumption is that the supersymmetries are
parallel in the unwarped metric\ft{This assumption is necessary in order to
compare the conditions on $G$ with those which follow from \cite{guk}. This is
because, \cite{guk} implicitly assumes that supersymmetric cycles in a
$G_2$-holonomy manifold are calibrated submanifolds. This assumption is valid
for supersymmetries which are parallel. }  , which
implies that $\eta$ is an eigenvector of the matrix
${11 \over 4} {\gamma}^p {\nabla}_p \log (\Delta )$ with eigenvalue
${3i \over 2} m {\Delta}^{3 \over 2}$. However, the eigenvalues of that matrix must
be real, so we obtain that
\be
{\nabla}_p \Delta = m = 0
\ee
ie the warp factor and Freund-Rubin parameter are trivial.

The spinor $\eta$
may be taken to be the product of a parallel spinor $\epsilon$
on $M^4$ and a spinor $\theta$ on $M^7$.
It follows from the work of Candelas and Raine
\cite{cr}
that the remaining constraints imposed by equations $(2)$
on such backgrounds
imply that
\be
{\nabla}_{n} \theta = m = G_{pqrs} = 0
\ee

These equations can be proven simply from integrability of $(2)$ with
warp factor one. We do not give the details here since we will be performing
a more general calculation below - of which the above is a special case.
The fact that $({M^7}, g_{7})$
admits a parallel spinor is equivalent
to the statement that the holonomy group of $g_7$ is $G_2$ or a subgroup
thereof.

Note that a slightly stronger result can be achieved without assuming
the supersymmetry is parallel but by assuming that the Freund-Rubin parameter
$m$ is zero. Then the integrability equations, derived from $(11)$ can be used
to show that when $M^7$ is compact the $G$-field vanishes, the warp factor
is trivial and that the spinor is in fact parallel. We will not give the details of
this here, since it implies the same conditions on the background spacetime as
above.

More generally we can require that $({M^4},{g_4})$ admit Killing spinors.
The most general such spinor obeys\ft{Tensor products are to be understood
throughout the remainder of the paper.},
\be
{\nabla}_{\mu}\epsilon = {\Lambda}_{1}{\gamma}_{\mu}\epsilon +
i{\Lambda}_{2}{\gamma}_{5}{\gamma}_{\mu}\epsilon
\ee
where the constants ${\Lambda}_{1}$ and ${\Lambda}_{2}$ are real\ft{Reality
follows from the Majorana condition on $\eta$ $\equiv$
${\epsilon} \otimes \theta$.}.
The case in which the right hand side of
$(13)$ is zero is essentially the case discussed above.

Here one again obtains an equation of the form of $(6)$ but with
$\alpha$ and $\beta$ given by
\be
\alpha \equiv {1 \over 144}G_{mnpq}{\gamma}^{mnpq} + {\Lambda}_1
\ee
\be
{\beta}_2 \equiv {\Lambda}_2 - m
\ee
The warp factor is assumed constant in the remainder of this section.
Contracting this new version of $(6)$ with ${\gamma}^{\mu}$ we find
\be
( \alpha -i {\beta}_2 {\gamma}_{5}) \eta = 0
\ee
from which it follows, since $\alpha$, ${\beta}_2$ and ${\gamma}_5$
are hermitian that
\be
\alpha \eta = {\beta}_2 \eta = 0
\ee

Next we consider the $n$-components of $(2)$. Taking our ansatz and
substituting $(19)$ we obtain
\be
{\nabla}_m \eta - {\Lambda}_{1} {\gamma}_{5} \otimes {\gamma}_m \eta
- {1 \over 12}{\gamma}_{5} \otimes G_{mpqr}{\gamma}^{pqr} \eta -
{i \over 2} {\Lambda}_2
\otimes {\gamma}_m \eta = 0
\ee

Contracting $(20)$ with ${\gamma}^m$ and operating
from the left with the Dirac operator, we find
\be
({{\nabla}_m}^2 + {1 \over 4} R) \eta - 25{\Lambda}_{1}^2 \eta
+ 35i{\Lambda}_{1}{\Lambda}_2 {\gamma}_{5} \eta
+ {49 \over 4} {\Lambda}_2^2 \eta = 0
\ee
This equation implies that
\be
{\Lambda}_{1}{\Lambda}_2 = 0
\ee

On the other hand, operating on $(20)$ with ${\nabla}_n$, taking the
skew-symmetric part and contracting the result with ${\gamma}^{mn}$,
we obtain after some algebra and use of identities found in the appendix
\be
{1 \over 4} R = -12 {\Lambda}_1^2 - {21 \over 2} {\Lambda}_2^2 -
{7 \over 48} G_{pqrs}^2
\ee

In deriving the above a hermiticity argument also yields that
\be
{\nabla}_{n}{G^n}_{pqr}{\gamma}^{pqr} \eta = 0
\ee

Combining equation $(23)$ with $(21)$ we obtain our main result:
\be
{\nabla}_m^2 \eta - {7 \over 48}G_{pqrs}^2 \eta - 37{\Lambda}_1^2
\eta +
{7 \over 4}{\Lambda}_2^2 \eta = 0
\ee

Recall $(22)$ that either ${\Lambda}_1$ or ${\Lambda}_2$ must be zero.
When ${\Lambda}_2$ is zero, the fact that ${\nabla}_m^2$ is negative
semi-definite implies that both $G$ and ${\Lambda}_1$ are also zero ie
our first solution is summarised as
\be
{\nabla}_m \eta = G_{pqrs} = m = {\Lambda}_1 = {\Lambda}_2 = 0
\ee
ie there is no flux and the pair $({M^7},{g_7})$ are a $G_2$-holonomy
7-manifold. This is the solution of \cite{cr} stated earlier.
Next we consider the case that ${\Lambda}_1$ is zero. In this case we have
not been able to find a solution in which $G_{pqrs}$ is non-zero either. For
instance, contraction of $(18)$ with ${\gamma}^m$ yields in this case
\be
{\gamma}^m {\nabla}_m \eta -{7i \over 2}{\Lambda}_2 \eta = 0
\ee

This equation is implied by
\be
{\nabla}_m \eta -{i \over 2}{\Lambda}_2  {\gamma}_m \eta = 0
\ee

which obviously solves our main equation with $G_{pqrs}$ $=$ $0$.
Thus, we obtain our second solution which is summarised by the following
three equations
\be
G_{pqrs} = {\Lambda}_1 = 0
\ee
\be
m = {\Lambda}_2
\ee
and
\be
{\nabla}_m \eta -{i \over 2}{\Lambda}_2  {\gamma}_m \eta = 0
\ee

This second solution is nothing but the original -well studied-
supersymmetric Freund-Rubin ansatz .

In summary thus far, we have considered the most general metric on
a spacetime of topology $X$ $=$ ${M^4}{\times}M^7$. We assumed that
the supersymmetry parameter is parallel in the unwarped metric. With this
assumption, supersymmetry requires that the 11-metric on $X$ is a Riemannian
product ie that the warp factor is trivial.

With trivial warp factor, for completeness, we included a derivation of
how supersymmetry then leads to two solution classes - a fact well known
from the Kaluza-Klein supergravity studies, see \cite{KKS} and references
therein.
Firstly when $({M^4}, {g_4})$
admits a parallel spinor, $({M^7}, {g_7})$ must do so as well and
there is no background $G$-field.
The second class of solutions are such that
that $({M^4}, {g_4})$ admits a non-trivial
Killing spinor, in which case $({M^7}, {g_7})$ does too. In this case there is
one component of $G$ which is non-zero and proportional to the
volume form of $M^4$.

\newpage
\Large
\noindent
{\bf {\sf 3. Relation to the Gukov Superpotential.}}
\normalsize
\bigskip

In \cite{guk} Gukov considered the relationship
between calibrated submanifolds
in $G_2$ holonomy manifolds, domain walls and the superpotential of the
four dimensional theory. The superpotential he conjectured for $M$ theory
compactified on a $G_2$ holonomy 7-manifold was obtained by extending
an argument of Gukov, Vafa and Witten who studied $M$ theory on
Calabi-Yau fourfolds. In \cite{gvw} it was observed that the conditions for
unbroken supersymmetry which follow from this conjectured form of the
superpotential agree in the fourfold case with the constraints imposed by
supersymmetry on the background four-form field strength which follow from
solving $(2)$ on fourfolds. This latter calculation was performed by Becker
and Becker. We will consider a similar comparison here.

In \cite{guk} Gukov proposed a superpotential for compactifications of
$M$ theory on a $G_2$ holonomy 7-manifold
$M^7$. Such a manifold admits a parallel $G_2$-
structure $\varphi$ (a locally $G_2$-invariant three form). The proposed
superpotential is given by
\be
W = {\int}_{M^7} G {\we} \varphi
\ee

As it stands this cannot be precisely correct since the right hand side is
manifestly real whereas in background Minkowski space the superpotential is a
holomorphic function of the chiral superfields\ft{Obviously, the value of the
superpotential in a vacuum could be real.}.
This follows from the fact
that  the effective four dimensional supergravity has four dimensional ${\cal
N}$ $=$ $1$ supersymmetry.
Recall \cite{bsa,hm} that the massless complex scalar
fields ${\Phi}_i$ in the low energy compactified theory are given by periods of
the ``complexified Bonan class'' over a basis ${\Sigma}_j$
of $b_{3}({M^7})$ cycles spanning $H_{3}({M^7},{\bf Z})$:
\be
{\Phi}_j = {\int}_{{\Sigma}_j}  i \varphi +  C
\ee
where $C$ is the locally defined $M$ theory three-form potential \ft{The
terminology ``complexified Bonan class'' stems from two facts: firstly that
Bonan discovered that $G_2$-holonomy 7-manifolds admit a parallel, locally
$G_2$ invariant three form \cite{bon}
and secondly the analogy with
``complexified Kahler class'' in string theory on Calabi-Yau manifolds.}

We should also consider the fact that $M$ theory compactified on a seven
manifold of $G_2$-holonomy can possess $M$2-brane domain walls which
reside at a point on the seven manifold. These are BPS saturated. Applying
the arguments of \cite{gvw,guk} to include the contribution from these walls,
one
must add a term in the superpotential proportional to ${\int}_{M^7} *G$
$+$  $G {\we} C $. The second term is another reason why the superpotential
is $(33)$ and not $(32)$.
Lorentz invariance implies that the first term ${\int}_{M^7} *G$ is
proportional to $mVol({M^7})$.

Combining these observations we see that
Gukov's  superpotential should properly be extended to \be
W({\Phi}_{j}) = m Vol({M^7}) + {\int}_{M^7} G {\we} ( i \varphi +  C )
\ee

The real and imaginary parts correspond respectively to $M$2-brane and
$M$5-brane domain walls in four dimensions. Note that the real part is
the Page charge.
In a four dimensional supergravity theory, the conditions for unbroken
supersymmetry and zero cosmological constant are given by solutions to
\be
W = dW = 0
\ee

Before we analyse these equations we must first make some remarks on
the cohomology groups of $G_2$-holonomy 7-manifolds. Since the holonomy
group is $G_2$, the spaces of tensor and spinor fields on $M^7$ decompose
into irreducible $G_2$ representations (or more specifically modules).
This decomposition, in the case of ${\Lambda}^{*}({TM^7})$ commutes with
the Laplacian and hence descends to the cohomology groups of $M^7$.
This gives the analogue \cite{J1}
of the Hodge-Dolbeaut cohomology groups of
a Kahler manifold. For instance, in the case of $H^{4}({M^7}, {\bf Z})$ we
have
\be
H^{4}({M^7}, {\bf Z}) = H^{4}_{1}({M^7}, {\bf Z}) \oplus H^{4}_{7}({M^7}, {\bf
Z}) \oplus H^{4}_{27}({M^7}, {\bf Z})
\ee
The subscripts denote the irreducible representations of $G_2$. Note that
the above corresponds to the decomposition of the space of four-forms,
which on a generic 7-manifold is a ${\bf 35}$ of $SO(7)$, under $G_2$:
\be
{\bf 35} \longrightarrow {\bf 1} + {\bf 7} + {\bf 27}
\ee

One should also note that Gukov's potential should really be regarded as
giving information about the cohomology class of $G$, since it was derived
by considering the charges of domain walls and these are determined by
cohomology classes. By abuse of notation below we will denote the
cohomology class of $G$-field by $G$.

The first condition in $(35)$ implies that
\be
G {\we} \varphi = 0
\ee

This means that the singlet piece of $G$ under $G_2$ vanishes.
We also have from the vanishing of the real part of $W$ that
\be
{\int} G{\we}C + m Vol({M^7}) = 0
\ee

which states that the Page charge vanishes. We will return to this condition
momentarily.

The variations of the real parts of the superfields correspond to variations
of $\varphi$ in response to changes in the metric tensor. As is well
known \cite{gpp,J1}, these generate $H^{3}_{27}({M^7}, {\bf Z})$. Hence, this
implies that the piece of $G$ transforming as a ${\bf 27}$ of $G_2$ is also
zero. Finally, for manifolds whose holonomy is strictly $G_2$ and not a
subgroup $H^{4}_{7}({M^7}, {\bf Z})$ is trivial \cite{J1}, since
such manifolds have finite fundamental group. Hence, (the cohomology class
of)
$G$ is identically zero in supersymmetric backgrounds
of low energy $M$ theory on $G_2$-holonomy 7-manifolds.
With $G$ trivial in cohomology, the first term in the Page charge
is zero and hence $m$ must also
be zero.

Thus, we have confirmed that Gukov's superpotential gives
the same answer as the equations which follow from eleven dimensional
supergravity.

For AdS space we have not been able to find a $G_2$ holonomy solution
so we cannot compare to Gukovs work.

\newpage

\Large
\noindent
{\bf {\sf 4. Discussion.}}
\normalsize
\bigskip

Thus far we have only discussed the eleven dimensional supergravity
approximation to $M$ theory. This receives quantum corrections which are
higher order in derivatives. In particular, consider the equation of motion
for the three-form potential. Schematically this has the form
\be
d*G = G{\we}G
\ee
in the eleven dimensional supergravity theory. Both our solutions satisfy
this equation.
However, in $M$ theory there
are corrections to this equation. To conclude, we will discuss these
corrections in a very speculative light.

Our first interest will be in the topologically
non-trivial corrections to this equation of which one term is known
\be
d*G = G{\we}G + X_8
\ee
where $X_8$ represents, when restricted to any eight dimensional submanifold,
${1 \over 24}$ times its Euler density.
In fact following the work of \cite{dmw} it is very tempting to believe that
$X_8$ is the only cohomologically non-trivial correction to the equation
of motion\ft{The
first author is indebted to G. Moore for discussions on this point.}.

At the order in the derivative expansion at which the $X_8$ correction
arises, apparently one cannot write down an action which is invariant
under the supersymmetry transformation rules of the eleven dimensional
supergravity theory \cite{kp}. Rather, the supersymmetry transformations
themselves need to be corrected. We can ask whether or not the background
field strength is zero in this corrected theory. In other words, is $G$ $=$
$0$ implied by supersymmetry at higher orders in the derivative expansion? If
the answer to that question is yes, then we arrive at an interesting
conclusion: the equation of motion above gives a topological constraint on
spacetime, since in that case,

\be
X_8 = 0
\ee
in cohomology.
For instance it
implies that the Euler number of any 8-dimensional submanifold of spacetime
is zero.
If we now consider the path integral for $M$ theory regarded in a low
energy approximation as a sum over spacetime topologies then we might
come across an eleven manifold ${M^4}{\times}M^7$ which admits parallel
spinors but does not satisfy the above equation. Thus for such a manifold,
we necessarily have to turn on $G$ to compensate for $X_8$ - but this breaks
supersymmetry with our assumptions. Thus spacetime
topology might in this sense break supersymmetry. In the appendix we
discuss examples of eleven manifolds with these properties.

%As a simple example of an eleven manifold satisfying these criteria consider
%a sector of the Euclidean path integral in which $M^4$ is a multi-center
%gravitational instanton and $M^7$ is a $G_2$-holonomy 7-manifold with finite
%but non-trivial fundamental group. Both types of manifolds exist. Then the
%product eleven-manifold admits 8-cycles of the form ${\bf S^2}{\times}{\bf
%N^6}$ where ${\bf N^6}$ is a 6-cycle in $M^7$ with non-zero Euler number.
%Another possible source of examples in the Lorentzian theory concerns
%four manifolds $({M^4},{g_4})$ which have non-trivial ${\bf R^2}$ holonomy, as
%discussed in \cite{jose}. One again requires examples with non-trivial
%two-cycles of non-zero Euler number. We do not know of any examples.

However, a much more likely scenario is that at next non-trivial order in the
derivative expansion, one can turn on $G$ whilst still maintaining
supersymmetry. One would also require a modification of the $G_2$ holonomy
metric. This scenario is much more in line with a similar discussion in the
context of strongly coupled heterotic string theory on Calabi-Yau manifolds
\cite{wit}. If a similar story exists at higher order there is also a possible
scenario for low energy supersymmetry breaking.

Assume that at higher order in derivatives
one has a supersymmetric solution of $M$ theory in which spacetime is
topologically a product ${M^4}{\times}M^7$ where the four manifold is
Minkowski space
and the 7-manifold
admits $G_2$-holonomy metrics, even
though the higher derivative solution does not have $G_2$-holonomy.
This spacetime has
trivial eighth de Rham cohomology group, so $X_8$ in this case is necessarily
trivial. Then it is conceivable that the solution might in fact obey the lowest
order equation of motion $(38)$ and that the higher derivative corrections
cancel amongst themselves. In such a scenario, a low energy
(ie eleven dimensional supergravity) observer would observe
that supersymmetry is in fact broken,
whereas a high energy ($M$ theory) observer would not. These speculations
are currently undergoing a more scientific investigation \cite{bbill}.

\bigskip

\Large
\noindent
{\bf {\sf Acknowledgements.}}
\normalsize
\bigskip

We would like to thank F. Dowker, J. Gauntlett, C. M. Hull and G. Moore
for discussions. We are also indebted to M. Rozali for some early discussions
on the subject of this paper and K. Peeters for describing some of the results
of \cite{kp} prior to arXiving. The first author would also like to thank PPARC
by whom his research has been supported.

\newpage

\Large
\noindent
{\bf {\sf 5. Appendix.}}
\normalsize
\bigskip

We follow the conventions of \cite{cr}. These are related to the
conventions of \cite{KKS} by multiplying the gamma-matrices by $i$.
The spacetime eleven manifold is denoted by $X$. This is always topologically
the product of a (Lorentzian) four manifold $M^4$ and a (Euclidean)
7-manifold $M^7$.

\bigskip
\noindent
$*$ $M$, $N$, etc denote eleven dimensional world or tangent space indices.

\noindent
$*$ $\mu$, $\nu$ etc. denote four dimensional indices.

\noindent
$*$ $m,n,p$ etc denote seven dimensional indices.

The local coordinates of $M^4$ are denoted collectively by $x$ whilst those
of $M^7$ are denoted by $y$.

The gamma-matrices ${\Gamma}_M$ are hermitian for $M$ = $1,..,10$ whereas
${\Gamma}_0$ is anti-hermitian.

They obey

\be
\{ {\Gamma}_M , {\Gamma}_N \} = 2 g_{MN}
\ee

where the metric has signature $(-,+,+,.....,+)$

A decomposition of the matrices ${\Gamma}_M$ into gamma-matrices
appropriate to $M^4$ and $M^7$ is

\be
{\Gamma}_{\mu} = {\gamma}_{\mu} \otimes {\mathbb{1}} , \;\;\; {\Gamma}_{m} =
{\gamma}_{5} \otimes {\gamma}_m
\ee
where
\be
{\gamma}_5 = {i \over 4!} {\epsilon}_{\mu\nu\rho\sigma}
{\gamma}^{\mu}{\gamma}^{\nu}{\gamma}^{\rho}{\gamma}^{\sigma}
\ee
where ${\gamma}_{5}$ squares to one.

Defining
\be
G_n \equiv G_{npqr}{\gamma}^{pqr}, \;\;\;\; G \equiv G_{pqrs} {\gamma}^{pqrs}
\ee

the identities we used in the calculation are:
\be
{\gamma}^m {\gamma}^n {G_m} G_n = -6 G_n^2 - G^2
\ee
\be
{\gamma}^{mn} G_m G_n = -7 G_n^2 - G^2
\ee
\be
{\gamma}^{mn} G_m {\gamma}_n = 0
\ee
\be
{\gamma}^{mn} {\gamma}_n G_m = 6G
\ee
\be
G_m^2 = -{1 \over 8} G^2 - 3 G_{pqrs}^2
\ee

As a simple example of an eleven manifold satisfying the criteria
required in the discussion of section four, consider
a sector of the Euclidean path integral in which $M^4$ is a multi-center
gravitational instanton and $M^7$ is a $G_2$-holonomy 7-manifold with finite
but non-trivial fundamental group. Both types of manifolds exist. Then the
product eleven-manifold admits 8-cycles of the form ${\bf S^2}{\times}{\bf
N^6}$ where ${\bf N^6}$ is a 6-cycle in $M^7$ with non-zero Euler number.
Another possible source of examples in the Lorentzian theory concerns
four manifolds $({M^4},{g_4})$ which have non-trivial ${\bf R^2}$ holonomy, as
discussed in \cite{jose}. One again requires examples with non-trivial
two-cycles of non-zero Euler number. We do not know of any examples of
this type.

\end{document}